\documentclass[proceedings,letterpaper]{JHEP3}

\catcode`\@=11
\renewcommand\speaker[1]{\if@speaker\global\@dblspeaktrue\fi
                        \global\@speakertrue
                        \global\setbox\@firstaubox
                        \hbox{\let\thanks\@gobble
                                \let\footnote\@gobble 
                                \rm #1}%
                        #1
                        }%
\catcode`@=12

\PrHEP{PrHEP hep2001}
\conference{International Europhysics Conference on HEP}

\usepackage{epsfig}

\def\omit#1{{}}
\def\lsim{\mathrel{\mathpalette\vereq<}\!\!}

\def\vereq#1#2{\lower3.5pt\vbox{\baselineskip1pt\lineskip1pt
    \ialign{\\$#1\hfill##\hfil\\$\crcr#2\crcr\sim\crcr}}}

\newcommand{\beq}{\begin{equation}}
\newcommand{\eeq}{\end{equation}}
\newcommand{\beqa}{\begin{eqnarray}}
\newcommand{\eeqa}{\end{eqnarray}}

\def\lqcd{\Lambda_{\rm QCD}}
\def\d{{\rm d}}
\def\FD{{\cal F}}
\def\FDs{\FD_*}
\def\FDt{\FD_{(*)}}
\def\B0bar{\overline{B}{}^0}
\def\D0bar{\overline{D}{}^0}
\def\K0bar{\overline{K}{}^0}
\def\gev{{\rm GeV}}
\def\mev{{\rm MeV}}

\title{New Results on Flavor Physics}

\author{\speaker{Zoltan Ligeti}\\
	Ernest Orlando Lawrence Berkeley National Laboratory\\
	University of California, Berkeley, CA 94720\\
	E-mail: \email{zligeti@lbl.gov}}

\abstract{Recent progress in flavor physics is discussed.  In particular, I
review theoretical and experimental developments relevant for semileptonic $B$
decays and the determination of $|V_{cb}|$ and $|V_{ub}|$, for exclusive rare
decays, for nonleptonic $b\to c$ decays and tests of factorization, and for $D$
meson mixing. \hfill {\footnotesize LBNL--49214}}

\begin{document}

\section{Introduction}

The goal of the $B$ physics program is to precisely test the flavor structure
of the standard model (SM), that is the Cabibbo-Kobayashi-Maskawa (CKM)
description of quark mixing and $CP$ violation.  In the last decade the
accuracy with which we know that gauge interactions are described by the SM
improved by an order of magnitude, and sometimes more.  In the coming years 
tests of the flavor sector of the SM and our ability to probe for flavor
physics and $CP$ violation beyond the SM will improve in a similar manner.  

However, in contrast to the hierarchy problem of electroweak symmetry breaking,
there is no similarly robust argument that new flavor physics must appear near
the electroweak scale.  Nevertheless, the flavor sector provides severe
constraints for model building, and many extensions of the SM do involve new
flavor physics which may be observable at the $B$ factories.  Flavor physics
also played an important role in the development of the SM: (i)~the smallness
of $K^0-\K0bar$ mixing led to the GIM mechanism and a calculation of the charm
mass before it was discovered; (ii)~$CP$ violation led to the proposal that
there should be three generations before any third generation fermions were
discovered; and (iii)~the large $B^0-\B0bar$ mixing was the first evidence for
a very large top quark mass.

The $B$ meson system has several features which makes it well-suited to study
flavor physics and $CP$ violation.  Because the top quark in loop diagrams is
neither GIM nor CKM suppressed, large $CP$ violating effects are possible, some
of which have clean interpretations.  For the same reason, a variety of rare
decays are expected to have large enough branching fractions to allow for
detailed studies.  Finally, some of the hadronic physics can be understood
model independently because $m_b \gg \lqcd$.

In the standard model all flavor changing processes are mediated by charged
current weak interactions, whose couplings to the six quarks are given by a
three-by-three unitary matrix, the Cabibbo-Kobayashi-Maskawa (CKM) matrix.  It
has a hierarchical structure, which is well exhibited in the Wolfenstein
parameterization,
\begin{equation}
V_{\rm CKM} = \pmatrix{ V_{ud} & V_{us} & V_{ub} \cr
  V_{cd} & V_{cs} & V_{cb} \cr
  V_{td} & V_{ts} & V_{tb} } 
= \pmatrix{ 1-\frac{1}{2}\lambda^2 & \lambda & A\lambda^3(\rho-i\eta) \cr
  -\lambda & 1-\frac{1}{2}\lambda^2 & A\lambda^2 \cr
  A\lambda^3(1-\rho-i\eta) & -A\lambda^2 & 1} + \ldots \,.
\end{equation}
This form is valid to order $\lambda^4$.  The small parameter is chosen as the
sine of the Cabibbo angle, $\lambda \simeq 0.22$, while $A$, $\rho$, and $\eta$
are order unity.  In the SM, the only source of $CP$ violation in flavor
physics is the phase of the CKM matrix, parameterized by $\eta$.  The 
unitarity of $V_{\rm CKM}$
\FIGURE{
\epsfig{file=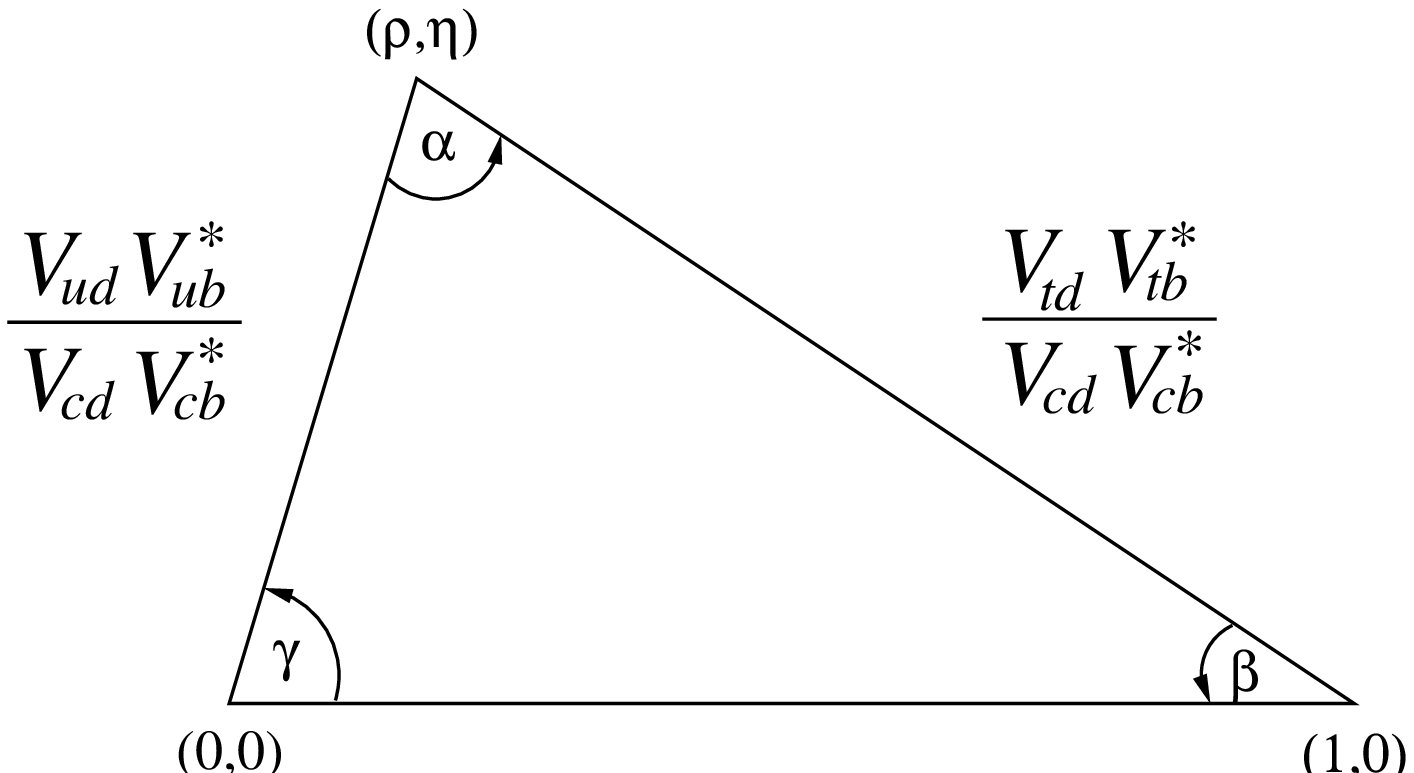, width=.39\textwidth}
\caption{The unitarity triangle}
\label{fig:triangle}
}
\noindent
implies that the nine complex elements of this matrix must satisfy $\sum_k
V_{ik} V_{jk}^* = \sum_k V_{ki} V_{kj}^* = \delta_{ij}$.\break The vanishing of
the product of the first and third columns provides a simple and useful way to
visualize  these constraints,
\begin{equation}
V_{ud}\, V_{ub}^* + V_{cd}\, V_{cb}^* + V_{td}\, V_{tb}^* = 0 \,,
\end{equation}
which can be represented as a triangle (see Fig.~\ref{fig:triangle}). Making
overconstraining measurements of the sides and angles of this unitarity
triangle is one of the best ways to look for new physics.

To believe at some point in the future that a discrepancy is a sign of new
physics, model independent predictions are essential.  Results which depend on
modeling nonperturbative strong interaction effects will never disprove the
Standard Model.  Most model independent  predictions are of the form
\begin{equation}
\mbox{Quantity of interest} = (\mbox{calculable factor}) 
  \times \bigg[ 1 + \sum_k\, (\mbox{small parameters})^k \bigg] \,,
\end{equation}
where the small parameter can be $m_s/\Lambda_{\chi SB}$, $\lqcd/m_b$,
$\alpha_s(m_b)$, etc.  Still, in most cases, there are theoretical
uncertainties suppressed by some $(\mbox{small parameter})^N$, which may be
hard to estimate model independently.  If one's goal is to test the Standard
Model, one must assign sizable uncertainties to such ``small"  corrections not
known from first principles.

Over the last decade, most of the theoretical progress in understanding $B$
decays utilized that $m_b$ is much larger than $\lqcd$.  However, depending on
the process under consideration, the relevant hadronic scale may or may not be
much smaller than $m_b$ (and, especially, $m_c$).  For example, $f_\pi$,
$m_\rho$, and $m_K^2/m_s$ are all of order $\lqcd$, but their numerical values
span more than an order of magnitude.  In many cases, as it will become clear
below, experimental guidance is needed to decide how well the theory works in
different cases.

To overconstrain the unitarity triangle, there are two very important ``clean"
measurements which will reach precisions at the few, or maybe even one, percent
level.  One is $\sin2\beta$ from the $CP$ asymmetry in $B\to J/\psi\, K_S$,
which is rapidly becoming the most precisely known ingredient of the unitarity
triangle~\cite{Gautier}.  The other is $|V_{td}/V_{ts}|$ from the ratio of the
neutral meson mass differences, $\Delta m_d/\Delta m_s$.  The LEP/SLD/CDF
combined limit is presently~\cite{Bsmixing} 
\beq 
\Delta m_s > 14.6\,{\rm ps} \quad (95\% \mbox{ CL})\,. 
\eeq 
Probably $B_s$ mixing will be discovered at the Tevatron, and soon thereafter
the experimental error of $\Delta m_s$ is expected to be below the 1\%
level~\cite{Barry}.  The uncertainty of $|V_{td}/V_{ts}|$ will then be
dominated by the error of $\xi \equiv (f_{B_s}/f_{B_d})
\sqrt{B_{B_s}/B_{B_d}}$.  For the last few years the lattice QCD averages have
been about $\xi = 1.15 \pm 0.06$~\cite{Sinead}, surprisingly consistent with
the chiral log calculation, $\xi^2 \sim 1.3$~\cite{Bslogs}.  This year we are
learning that an additional error, estimated to be
$^{+0.07}_{-0.}$~\cite{Sinead}, may have to be added to $\xi$ for now, since in
the unquenched calculation chiral logs are important in the chiral
extrapolation for $f_B$, but they do not affect $f_{B_s}$~\cite{JLQCD}.  It is
very important to reduce this uncertainty, and do simulations with three light
flavors.

Compared to $\sin2\beta$ and $|V_{td}/V_{ts}|$, for which both the theory and
the experiment are tractable, much harder is the determination of another side
or another angle, such as $|V_{ub}|$, or $\alpha$, or $\gamma$ ($|V_{cb}|$ is
also ``easy" by these criteria).  However, our ability to test the CKM
hypothesis in $B$ decays will depend on a third best measurement besides
$\sin2\beta$ and $x_s$ (and on ``null observables"). The accuracy of these
measurements will determine the sensitivity to new physics, and  the precision
with which the SM is tested.  It does not matter whether it is a side or an
angle.  What is important is which  measurements can be made that have clean
theoretical interpretations for the short distance physics we are after.

Section~\ref{sec:sl} reviews recent progress for semileptonic decays and the
determination of $|V_{cb}|$ and $|V_{ub}|$.  Related developments relevant for
exclusive rare decays are also discussed.  Section~\ref{sec:nl} deals with
nonleptonic decays, such as lifetimes, tests of factorization for exclusive
nonleptonic decay, and $D^0$ mixing.  Section~\ref{sec:concl} contains our
conclusions. While this write-up follows closely the slides at the Conference,
I attempted to update the experimental results where available.  I was asked
not to talk about $CP$ violation, which was reviewed in
Refs.~\cite{Gautier,Beneke}.

\section{Semileptonic decays}\label{sec:sl}

\FIGURE{
\epsfig{file=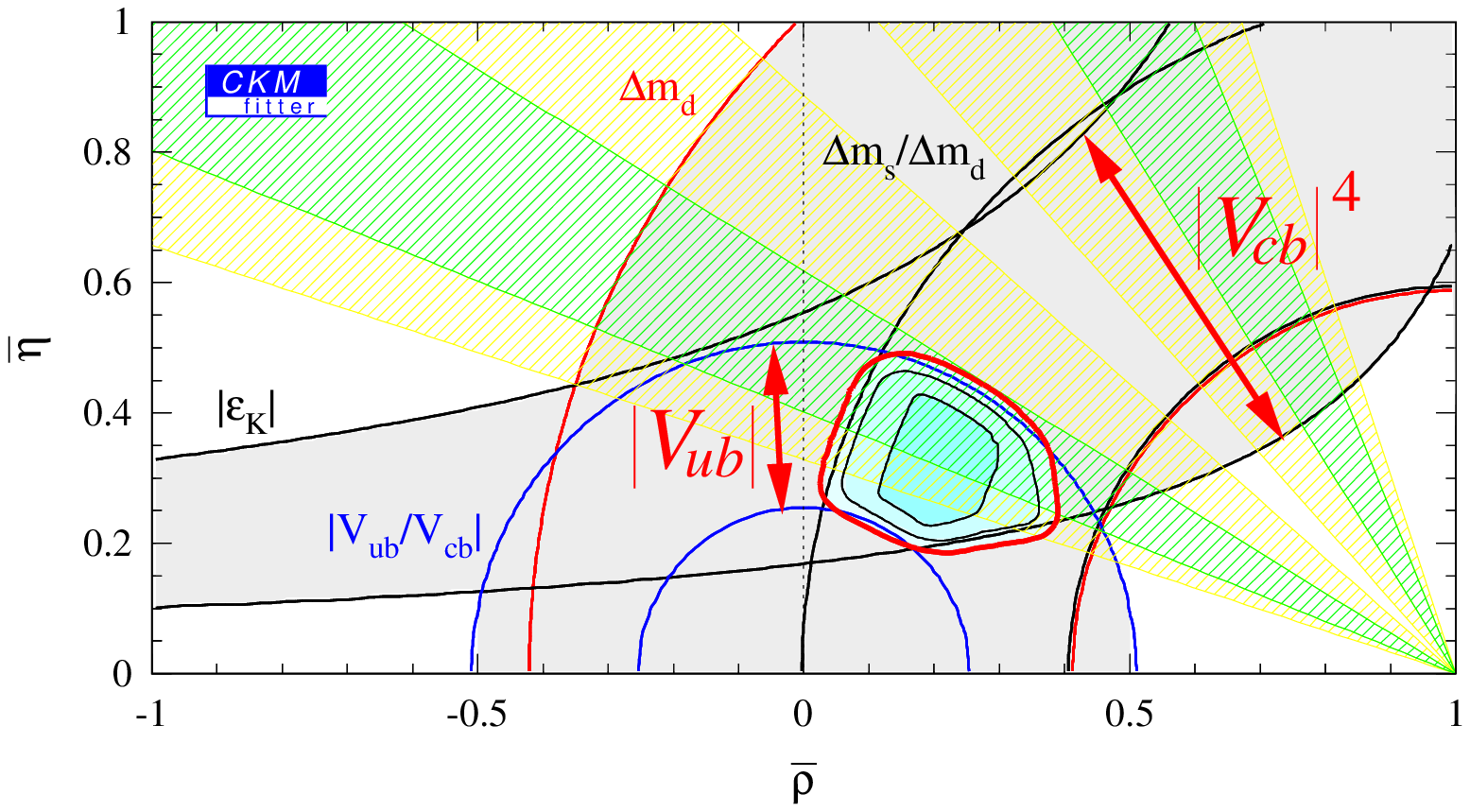, width=.47\textwidth}
\caption{The allowed range of $\bar\rho-\bar\eta$~\cite{ckmfitter}.}
\label{fig:ckmfitter}
}

The determination of $|V_{cb}|$ and $|V_{ub}|$ are very important for testing
the CKM hypothesis.  The allowed range of $\sin 2\beta$ in the SM depends
strongly on the uncertainty of $|V_{ub}|$ (since it determines the side of the
unitarity triangle opposite to the angle $\beta$), and the constraint from the
$K^0 - \K0bar$ mixing parameter $\epsilon_K$ is proportional to $|V_{cb}|^4$. 
This is illustrated in Fig.~\ref{fig:ckmfitter}.   Moreover, the methods
developed to extract $|V_{cb}|$ and $|V_{ub}|$ are also useful for reducing the
hadronic uncertainties in rare decays.

\subsection{\boldmath Exclusive $B\to D^{(*)}\ell\bar\nu$ decay and the HQET}

In heavy mesons composed of a heavy quark, $Q$, and a light antiquark, $\bar
q$, and gluons and $q\bar q$ pairs, in the $m_Q \to \infty$ limit the heavy
quark acts as a static color source with fixed four-velocity $v^\mu$.  The
wave-function of the light degrees of freedom become insensitive to the spin
and mass (flavor) of the heavy quark, resulting in heavy quark spin-flavor
symmetries~\cite{HQS}.

The determination of $|V_{cb}|$ from exclusive $B\to D^{(*)}\ell\bar\nu$ decays
is based on the fact that heavy quark symmetry relates the form factors which
occur in these decays to the Isgur-Wise function, whose value is known at zero
recoil in the infinite mass limit. The symmetry breaking corrections can be
organized in a simultaneous expansion in $\alpha_s(m_Q)$ and $\lqcd/m_Q$ (where
$Q = c,b$).  The $B\to D^{(*)}\ell\bar\nu$ rates can be schematically
written as
\begin{equation}\label{rates}
{\d\Gamma(B\to D^{(*)} \ell\bar\nu)\over \d w} = \mbox{(known factors)}\,
|V_{cb}|^2 \cases{\! (w^2-1)^{1/2}\, \FDs^2(w)\,,\quad & for $B\to D^*$, \cr
\!(w^2-1)^{3/2}\, \FD^2(w)\,, & for $B\to D$, \cr}
\end{equation}
where $w = (m_B^2 + m_{D^{(*)}}^2 - q^2) / (2m_B m_{D^{(*)}})$.  Both $\FD(w)$
and $\FDs(w)$ are equal to the Isgur-Wise function in the $m_Q \to\infty$
limit, and in particular $\FD(1) = \FDs(1) = 1$, allowing for a model
independent determination of $|V_{cb}|$. The zero recoil limits of $\FDt(w)$
are of the form
\begin{equation}\label{F1}
\FDs(1) = 1 + c_A(\alpha_s) + {0\over m_Q} 
  + {(\ldots)\over m_Q^2} + \ldots \,, \qquad
\FD(1) = 1 + c_V(\alpha_s) + {(\ldots)\over m_Q} 
  + {(\ldots)\over m_Q^2} + \ldots \,.
\end{equation}
The perturbative corrections, $c_A = -0.04$ and $c_V = 0.02$, have been
computed to order $\alpha_s^2$~\cite{Czar}, and the unknown higher  order
corrections should be below the 1\% level.  The  order $\lqcd/m_Q$  correction
to $\FDs(1)$ vanishes due to Luke's theorem~\cite{Luke}.  The terms  indicated
by $(\ldots)$ in Eqs.~(\ref{F1}) are only known using phenomenological  models
or quenched lattice QCD at present.  This is why the determination of
$|V_{cb}|$ from $B\to D^* \ell\bar\nu$ is theoretically more reliable for now
than that from $B\to D\ell\bar\nu$, although both QCD sum rules~\cite{LNN}
and quenched lattice QCD~\cite{latticeD} suggest that the order $\lqcd/m_Q$
correction to $\FD(1)$ is small.  Due to the extra $w^2-1$ helicity
suppression near zero recoil, $B\to D\ell\bar\nu$ is also harder
experimentally.

$|V_{cb}|\, {\cal F}_*(1)$ is measured from the zero recoil limit of the decay
rate, and the results are shown in Table~\ref{tab:Vcb}. The main theoretical
uncertainties in such a determination of $|V_{cb}|$ come
\TABLE{
\begin{tabular}{cc} \hline
$|V_{cb}|\, {\cal F}_*(1) \times 10^3$  &  Experiment \\ \hline\hline
$35.6 \pm 1.7$  &  LEP~\cite{Palla} \\
$42.2 \pm 2.2$  &  CLEO~\cite{CleoDs} \\
$36.2 \pm 2.3$  &  BELLE~\cite{BelleDs} \\ \hline
\end{tabular}
\caption{Measurements of $|V_{cb}|\, {\cal F}_*(1)$.}
\label{tab:Vcb}
}\noindent
from the value of $\FDt(w)$ at $w=1$ and from its shape  used to fit the data. 
In my opinion, a reasonable estimate at present is
\begin{equation}
{\cal F}_*(1) = 0.91 \pm 0.04\,,
\end{equation}
where the error can probably only be reduced by unquenched lattice calculations
in the future.  The quenched result is ${\cal F}_*(1) = 0.913 ^{+0.024 + 0.017}
_{-0.017 - 0.030}$~\cite{latticeDs}.  It will be interesting to see the effect
of unquenching, and whether $|V_{cb}|$ obtained from $B\to D\ell\bar\nu$ using
${\cal F}(1)$ from the lattice will agree at the few percent level.

For the shape of $\FDs(w)$, it is customary to expand about zero recoil and
write $\FDs(w) = \FDs(1)\, [1 - \rho^2\, (w-1) + c\, (w-1)^2 + \ldots]$.
Analyticity imposes stringent constraints between the slope, $\rho^2$, and
curvature, $c$, at zero recoil~\cite{BGL,CLN}, which is already used to fit the
data and obtain the results in Table~\ref{tab:Vcb}.  Recently there has been
renewed effort in constraining the slope parameter $\rho^2$ using sum rules and
data on $B$ decays to excited $D$ states~\cite{Oliver,Uraltsevsr}.  Decays to
orbitally excited $D$ mesons can also be studied in HQET~\cite{Falk,LLSW}, and
it seems problematic to accommodate the data which suggests that the rate to
the $D_1^*$ and $D_0^*$ states ($s_l^{\pi_l} = \frac12^+$) is larger than that
to $D_1$ and $D_2^*$ ($s_l^{\pi_l} =
\frac32^+$)~\cite{Oliver,Uraltsevsr,LLSW}.  

Measuring the $B\to D\ell\bar\nu$ rate~\cite{CleoD,BelleD} is also important,
since computing $\FD(1)$ on the lattice is no harder that $\FDs(1)$, and so it
provides an independent determination of $|V_{cb}|$.  Comparing the shapes of
the $B\to D^*$ and $B\to D$ spectra may also help, since it gives additional
constraints on $\rho^2$, and the correlation between $\rho^2$ and the extracted
value of $|V_{cb}|\, {\cal F}_*(1)$ is very large~\cite{BGZL}.

\subsection{\boldmath Inclusive semileptonic $B$ decay and the OPE}

Inclusive $B$ decay rates can be computed model independently in a series in
$\lqcd/m_b$ and $\alpha_s(m_b)$, using an operator product expansion
(OPE)~\cite{OPE}.  The results can be schematically written as
\begin{equation}\label{incl}
\d\Gamma = \pmatrix{b{\rm ~quark} \cr {\rm decay}\cr} \times 
\bigg\{ 1 + \frac0{m_b} + \frac{f(\lambda_1,\lambda_2)}{m_B^2} + \ldots
  + \alpha_s(\ldots) + \alpha_s^2(\ldots) + \ldots \bigg\} \,.
\end{equation}
The $m_b\to\infty$ limit is given by $b$ quark decay, and the leading
nonperturbative corrections suppressed by $\lqcd^2/m_b^2$ are parameterized by
two hadronic matrix elements, usually denoted by $\lambda_1$ and $\lambda_2$. 
The value $\lambda_2 \simeq 0.12\, {\rm GeV}^2$ is known from the $B^*-B$ mass
splitting.  At order $\lqcd^3/m_b^3$ seven new and unknown hadronic matrix
elements enter, and usually naive dimensional analysis is used to estimate
their size and the related uncertainty.  For most quantities of interest, the
perturbation series are known including the $\alpha_s$ and $\alpha_s^2\beta_0$
terms, where $\beta_0 = 11 - 2n_f/3$ is the first coefficient of the
$\beta$-function (which is large, so in many cases this term is expected to
dominate the $\alpha_s^2$ corrections). 

The good news from the above is that ``sufficiently inclusive" quantities, such
as the total semileptonic width relevant for the determination of $|V_{cb}|$,
can be computed with errors at the $\,\lsim 5\%$ level.  In such cases the
theoretical uncertainty is controlled dominantly by the error of a short
distance $b$ quark mass (whatever way it is defined).  Using the ``upsilon
expansion"~\cite{upsexp} the relation between the inclusive semileptonic rate
and $|V_{cb}|$ is
\begin{equation}\label{Vcb}
|V_{cb}| = (41.9 \pm 0.8_{(\rm pert)} \pm 0.5_{(m_b)} \pm 0.7_{(\lambda_1)}) 
  \times 10^{-3}\, \bigg( {{\cal B}(\bar B\to X_c \ell\bar\nu)\over0.105}\,
  {1.6\,{\rm ps}\over\tau_B} \bigg)^{1/2} \,.
\end{equation}
The first error is from the uncertainty in the perturbation series, the second
one from the $b$ quark mass, $m_b^{1S} = 4.73 \pm 0.05\,$GeV (a conservative
range of $m_b$ may be larger~\cite{bmasses}), and the third one from $\lambda_1
= -0.25 \pm 0.25\,{\rm GeV}^2$.  This result is in agreement with
Ref.~\cite{BSUreview}, where the central value is $40.8 \times 10^{-3}$
(including the 1.007 electromagnetic radiative correction).

LEP and BELLE reported new results for the semileptonic branching ratio, which
yield a determination of $|V_{cb}|$ which is dominated by theoretical errors,
\begin{equation}\label{Vcbincl}
{\cal B}(B\to X\ell\bar\nu) = \cases{ 
  10.65 \pm 0.23 \%  & (LEP~\cite{Palla}) \cr
  10.86 \pm 0.49 \%  &  (BELLE~\cite{Won}) \cr } \quad \Rightarrow\;
|V_{cb}| = (41 \pm 2_{(\rm th)}) \times 10^{-3}\,.
\end{equation}
Future improvements are likely to come from combined analyses using inclusive
spectra to determine $m_b$ and $\lambda_1$ (or, equivalently, $\bar\Lambda$ and
$\lambda_1$).  It had been suggested that moments of the $B\to X_c \ell\bar\nu$
lepton spectrum~\cite{Voloshin94,gremmetal,GK} or hadronic invariant mass
spectrum \cite{FLSmass,GK}, or the $B\to X_s\gamma$ photon
spectrum~\cite{KL,LLMW} can be used to determine these parameters.  Each
measurement is a band in the $\bar\Lambda - \lambda_1$ plane, and the
combination of several of them can pin down $\bar\Lambda$ and $\lambda_1$, and
also test theoretical assumptions of the method.  I.e., if  quark-hadron
duality were violated at the several percent level, it should show up as an
inconsistency.  

\FIGURE{
\epsfig{file=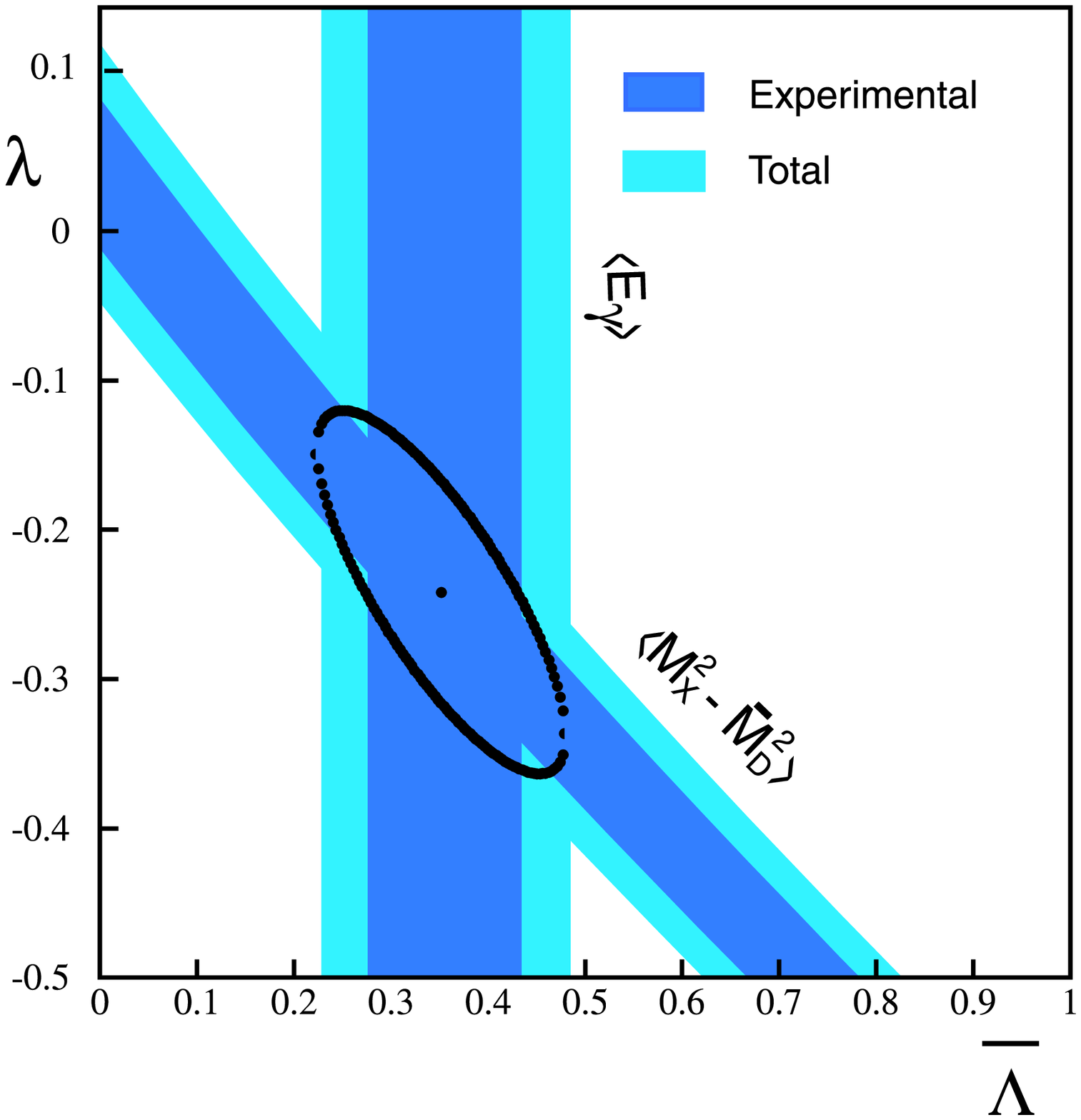, width=0.42\textwidth}
\caption{$\bar\Lambda$ and $\lambda_1$ from $\langle m_X^2 -
\overline{m}_D^2\rangle$ in $B\to X_c \ell\bar\nu$ and $\langle
E_\gamma\rangle$ in $B\to X_s\gamma$~\cite{CLEOhadmom}.}
\label{fig:moments}
}

The first such analysis was done recently by CLEO~\cite{CLEOhadmom,CLEObsgmom},
using the two moments shown in Fig.~\ref{fig:moments}.  Combining with their
semileptonic rate measurement, they obtain
\begin{equation}\label{Vcbmom}
|V_{cb}| = (40.4 \pm 1.3) \times 10^{-3}\,.
\end{equation}
The advantage of this measurement is that a sizable part of the
hard-to-quantify theory error in Eq.~(\ref{Vcbincl}) is traded for experimental
errors on the moment measurements.  To make further progress, one must quantify
better the accuracy of quark-hadron duality, but if no problems are encountered
$\sigma(|V_{cb}|) \sim 2\%$ may be achievable.

It will continue to be important to pursue both the inclusive and exclusive
measurements of $|V_{cb}|$.  Since both the theoretical and the experimental
systematic uncertainties are different, agreement between the two
determinations will remain to be a very powerful cross-check that the errors are
as well understood as claimed.

\subsection{\boldmath Inclusive $B\to X_u\ell\bar\nu$ spectra and $|V_{ub}|$}

If it were not for the $\sim\! 100$ times larger $b\to c$ background, measuring
$|V_{ub}|$ would be as ``easy" as $|V_{cb}|$.  The total $B\to X_u \ell\bar\nu$
rate can be predicted in the OPE with small uncertainty~\cite{upsexp},
\begin{equation}\label{Vub}
|V_{ub}| = (3.04 \pm 0.06_{(\rm pert)} \pm 0.08_{(m_b)}) \times 10^{-3}\,
  \bigg( {{\cal B}(\bar B\to X_u \ell\bar\nu)\over 0.001}
  {1.6\,{\rm ps}\over\tau_B} \bigg)^{1/2} \,,
\end{equation}
where the errors are as discussed after Eq.~(\ref{Vcb}).  The central value in
Ref.~\cite{BSUreview}, $3.24 \times 10^{-3}$, was later updated to $3.08 \times
10^{-3}$~\cite{Uraltsev99}.  If this fully inclusive rate is measured without
significant cuts on the phase space, then $|V_{ub}|$ can be determined with
small theoretical error.  It seems that measuring this rate fully inclusively
may become possible using the huge data sets expected in a couple of years at
the $B$ factories.

LEP reported measurements of the inclusive rate already, giving ${\cal B}(b\to
u \ell\bar\nu) = (1.71 \pm 0.31 \pm 0.37 \pm 0.21) \times
10^{-3}$~\cite{Palla}.  It is very hard from the outside to understand what
region of the Dalitz plot these results are sensitive to.  If it is the low
$X_u$ invariant mass region, then there is a sizable theoretical uncertainty
(see below).  As also emphasized in Refs.~\cite{HQhf9,MBWlp}, it would be most
desirable to present the results also in a form which is as theory-independent
as possible, and quote the rate as measured in a given kinematic region.

When kinematic cuts are used to distinguish the $b\to c$ background from the
$b\to u$ signal, the behavior of the OPE can be affected dramatically.  There
are three qualitatively different regions of phase space, depending on the
allowed invariant mass and energy (in the $B$ rest frame) of the hadronic final
state:

(i) $m_X^2 \gg E_X \lqcd \gg \lqcd^2$: the OPE converges, and the first few
terms are expected to give reliable result.  This is the case for the  $B\to
X_c\ell\bar\nu$ width relevant for measuring $|V_{cb}|$.

(ii) $m_X^2 \sim E_X \lqcd \gg \lqcd^2$: an infinite set of equally important
terms in the OPE must be resummed; the OPE becomes a twist expansion and
nonperturbative input is needed.

(iii) $m_X \sim \lqcd$: the final state is dominated by resonances, and it is
not known how to compute any inclusive quantity reliably.

\FIGURE{
\epsfig{file=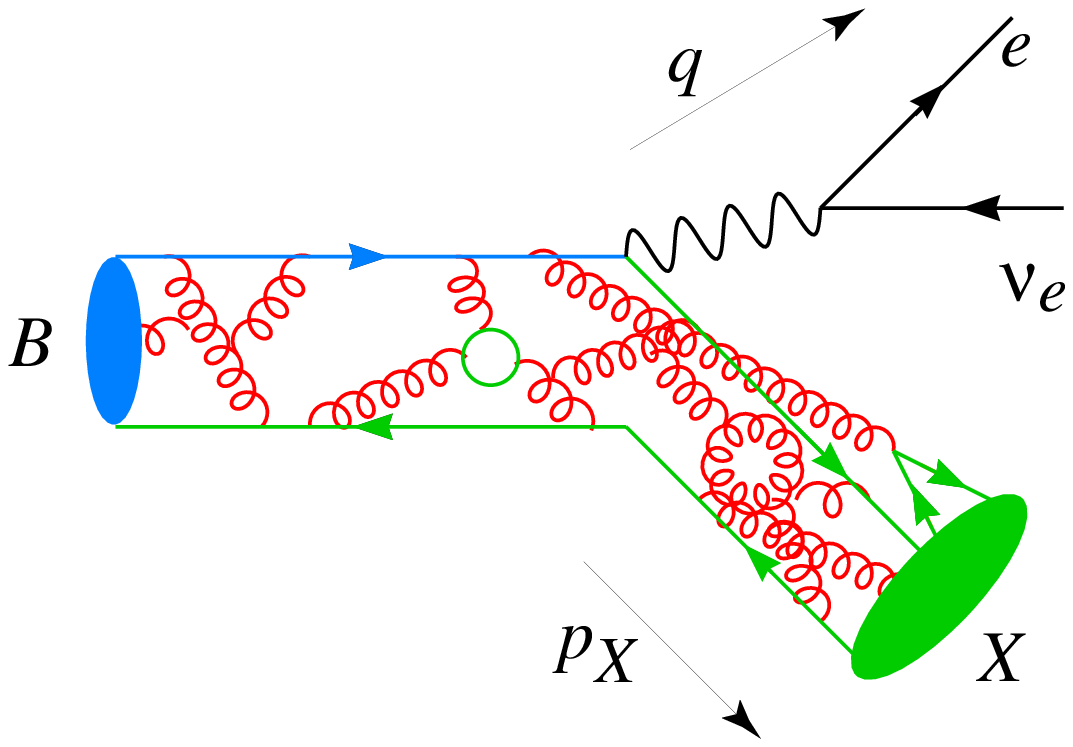, width=0.33\textwidth}
\caption{$B\to X\ell\bar\nu$ decay.}
\label{fig:mike}
}

Experimentally, there are several possibilities to remove the charm background:
the charged lepton endpoint region used to first observe $b\to u$ transition,
$E_\ell > (m_B^2-m_D^2) / (2m_B)$, the low hadronic invariant mass region, $m_X
< m_D$~\cite{FLW,BDU}, and the large dilepton invariant mass region $q^2 \equiv
(p_\ell + p_\nu)^2 > (m_B - m_D)^2$~\cite{BLL1}.  These contain roughly 10\%,
80\%, and 20\% of the rate, respectively.  Measuring $m_X$ or $q^2$ require
reconstruction of the neutrino, which is challenging.  

The problem for theory is that the phase space regions $E_\ell > (m_B^2-m_D^2)
/ (2m_B)$ and $m_X < m_D$ both belong to the regime (ii), because these cuts
impose $m_X \lsim m_D$ and $E_X \lsim m_B$, and numerically $\lqcd\, m_B \sim
m_D^2$.  The region $m_X < m_D$ is better than $E_\ell > (m_B^2-m_D^2) /
(2m_B)$ inasmuch as the expected rate is a lot larger, and the inclusive
description is expected to hold better.  But nonperturbative input is needed,
formally, at the ${\cal O}(1)$ level in both cases, which is why the model
dependence increases rapidly if the $m_X$ cut is lowered below
$m_D$~\cite{FLW}.  These regions of the Dalitz plot are shown in
Fig.~\ref{fig:dalitz}.

\FIGURE{
\epsfig{file=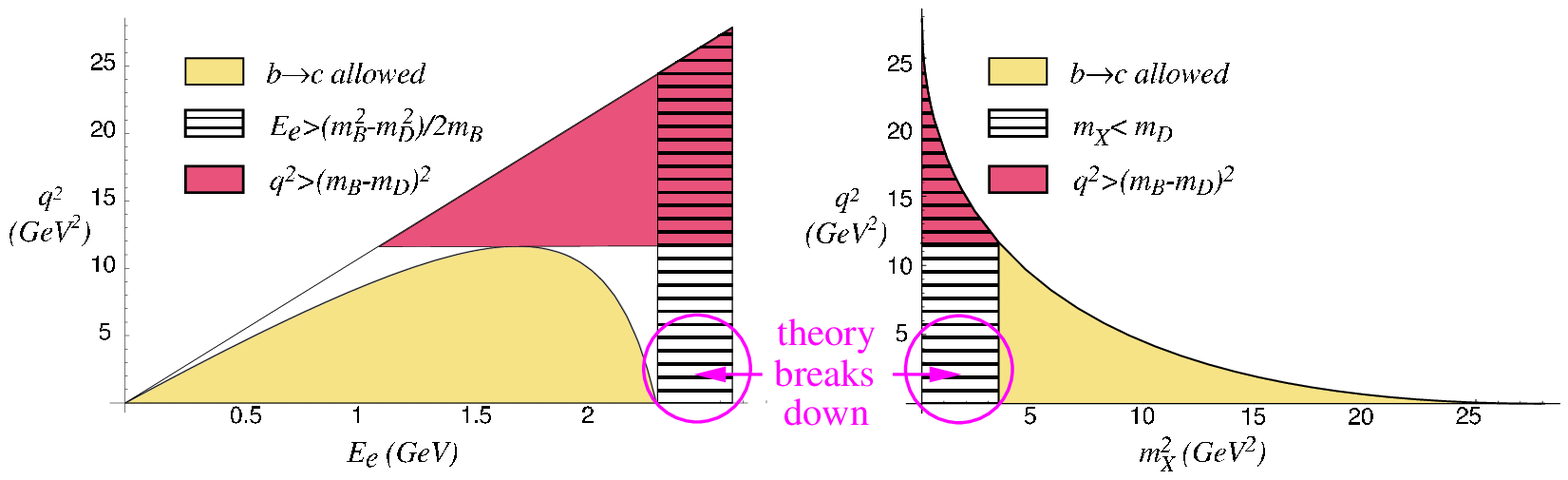, width=.95\textwidth}
\caption{Dalitz plots for $B\to X\ell\bar\nu$ decay in terms of $E_\ell$ and
$q^2$ (left), and $m_X^2$ and $q^2$ (right).}
\label{fig:dalitz}
}

The nonperturbative input needed to predict the spectra in the large $E_\ell$
and small $m_X$ regions, the $b$ quark light-cone distribution function
(sometimes also called shape function), is universal at leading order, and can
be related to the $B\to X_s\gamma$ photon spectrum~\cite{structure}. Recently
these relations have been extended to the resummed next-to-leading order
corrections, and applied to the large $E_\ell$ and small $m_X$
regions~\cite{extractshape}.  Weighted integrals of the $B\to X_s\gamma$ photon
spectrum are equal to the $B\to X_u\ell\bar\nu$ rate in the large $E_\ell$ or
small $m_X$ regions.  There is also a sizable correction from operators other
than $O_7$ contributing to $B\to X_s\gamma$~\cite{notO7}.  The dominant
theoretical uncertainty in these determinations of $|V_{ub}|$ are from
subleading twist contributions, which are not related to $B\to X_s\gamma$. 
These are suppressed by $\lqcd/m_b$, but their size is hard to quantify, and
even formulating them is nontrivial~\cite{subltwist}.  Of course, if the lepton
endpoint region is found to be dominated by the $\pi$ and $\rho$ exclusive
channels, then the applicability of the inclusive description may be
questioned.

In contrast to the above, in the $q^2 > (m_B-m_D)^2$ region the first few terms
in the OPE dominate~\cite{BLL1}.  This cut implies $E_X \lsim m_D$ and $m_X
\lsim m_D$, and so the $m_X^2 \gg E_X \lqcd \gg \lqcd^2$ criterion of regime
(i) is satisfied.  This relies, however, on $m_c \gg \lqcd$, and so the OPE is
effectively an expansion in $\lqcd/m_c$~\cite{neubertq2}.  The largest
uncertainties come from order $\lqcd^3/m_{c,b}^3$ nonperturbative corrections,
the $b$ quark mass, and the perturbation series.  Weak annihilation (WA)
suppressed by $\lqcd^3/m_b^3$ is important, because it enters the rate as
$\delta(q^2-m_b^2)$~\cite{Voloshin}.  Its magnitude is hard to estimate, as it
is proportional to the difference of two matrix elements of 4-quark operators,
which vanishes in the vacuum insertion approximation.  WA could be $\sim 2\%$
of the $B\to X_u\ell\bar\nu$ rate, and, in turn, $\sim 10\%$ of the rate in the
$q^2 > (m_B-m_D)^2$ region.  It is even more important for the lepton endpoint
region, since it is also proportional to $\delta(E_\ell - m_b/2)$.  Preliminary
lattice results of the matrix elements suggest a smaller size~\cite{MDP}. 
Experimentally, WA can be constrained by comparing $|V_{ub}|$ measured from
$B^0$ and $B^\pm$ decays, and by comparing the $D^0$ and $D_s$ semileptonic
widths~\cite{Voloshin}.

\TABLE{
\begin{tabular}{c||c|c} \hline
Cuts on  &  Fraction  &  Error of $|V_{ub}|$ \\
$q^2$ and $m_X$  &  of events  &  $\!\delta m_b=80/30\,\mev\!$ \\ \hline\hline
$6\,\gev^2,\, m_D$		& $46\%$ & $8\%/5\%$  \\
$8\,\gev^2,\, 1.7\,\gev$ 	& $33\%$ & $9\%/6\%$  \\ \hline
$\!(m_B-m_D)^2, m_D$ 		& $17\%$ & $15\%/12\%$ \\ \hline
\end{tabular}
\caption{$|V_{ub}|$ from combined cuts on $q^2$ and $m_X$.}
\label{tab:dblcut}
}

Combining the $q^2$ and $m_X$ cuts can significantly reduce the theoretical
uncertainties~\cite{BLL2}.  The right-hand side of Fig.~\ref{fig:dalitz} shows
that the $q^2$ cut can be lowered below $(m_B-m_D)^2$ by imposing an additional
cut on $m_X$.  This changes the expansion parameter from $\lqcd/m_c$ to
$m_b\lqcd/(m_b^2-q_{\rm cut}^2)$, resulting in a significant decrease of the
uncertainties from both the perturbation series and from the nonperturbative
corrections.  At the same time the uncertainty from the $b$ quark light-cone
distribution function only turns on slowly.  Some representative results are
give in Table~\ref{tab:dblcut}, showing that it is possible to determine
$|V_{ub}|$ with a theoretical error at the $5-10\%$ level using up to $\sim
45\%$ of the semileptonic decays~\cite{BLL2}.

\subsection{\boldmath Rare $B$ decays}

Rare $B$ decays are very sensitive probes of new physics.  There are many
interesting modes sensitive to different extensions of the Standard Model.  For
example, $B\to X_s\gamma$ provides the best bound on the  charged Higgs mass in
type-II two Higgs doublet model, and also constrains the parameter space of
SUSY models.  The photon spectrum, which is not sensitive to new physics, is
important for determinations of $|V_{ub}|$ and the $b$ quark mass, as discussed
earlier.  Other rare decays such as $B\to X\ell^+\ell^-$ are sensitive through
the $bsZ$ effective coupling to SUSY and left-right symmetric models. $B\to
X\nu\bar\nu$ can probe models containing unconstrained couplings between three
3rd generation fermions~\cite{GLN}.  In the Standard Model these decays are
sensitive to CKM angles involving the top quark, complementary to $B_{d,s}$
mixing.

This last year we learned that the CKM contributions to rare decays are likely
to be the dominant ones, as they probably are for $CP$ violation in $B\to \psi
K_S$.  This is supported by the measurement of ${\cal B}(B\to X_s\gamma)$ which
agrees with the SM at the 15\% level~\cite{CLEObsgmom}; the measurement of
$B\to K\ell^+\ell^-$ which is in the ballpark of the SM
expectation~\cite{Abe:2001dh}; and the non-observation of direct $CP$ violation
in $b\to s\gamma$, $-0.27 < A_{CP}(B\to X_s\gamma) < 0.10$~\cite{Coan:2000pu}
and $-0.17 < A_{CP}(B\to K^*\gamma) < 0.08$~\cite{Aubert:2001me} at the
90\%\,CL, which is expected to be tiny in the SM.  These results make it less
likely that we will observe orders-of-magnitude enhancements of rare $B$
decays.  It is more likely that only a broad set of precision measurements will
be able to find signals of new physics.

\TABLE{
\begin{tabular}{ccc} \hline
Decay  &  Approximate  &   Present  \\
mode  &  SM rate  &  status  \\ \hline\hline
$B\to X_s\gamma$  &  $3.5\times 10^{-4}$ & $(3.2 \pm 0.5)\!\times\! 10^{-4}$ \\
$B\to X_s\nu\bar\nu$  &  $4\times 10^{-5}$  &  $<7.7\times10^{-4}$  \\
$B\to \tau\nu$  &  $4\times 10^{-5}$  &  $<5.7\times10^{-4}$  \\
$B\to X_s \ell^+ \ell^-$  &  $7\times 10^{-6}$  &  $<1.0\times10^{-5}$  \\
$B_s\to \tau^+\tau^-$  &  $1\times 10^{-6}$  &  \\
$B\to X_s\tau^+\tau^-$  &  $5\times 10^{-7}$  &   \\
$B\to \mu\nu$  &  $2\times 10^{-7}$  &  $<6.5\times10^{-6}$  \\
$B_s\to \mu^+\mu^-$  &  $4\times 10^{-9}$  &  $<2\times10^{-6}$  \\
$B\to \mu^+\mu^-$  &  $1\times 10^{-10}$  &  $<2.8\times10^{-7}$ \\ \hline
\end{tabular}
\caption{Some interesting rare decays.}
\label{tab:rare}
}

At present, inclusive rare decays are theoretically cleaner than the exclusive
ones, since they are calculable in an OPE and precise multi-loop results
exist.  Table~\ref{tab:rare} summarizes some of the most interesting modes. 
The $b\to d$ rates are expected to be about a factor of $|V_{td}/V_{ts}|^2 \sim
\lambda^2$ smaller than the corresponding $b\to s$ modes shown.  As a
guesstimate, in $b\to q\, l_1 l_2$ decays one expects $10-20\%$ $K^*/\rho$ and
$5-10\%$ $K/\pi$.  A clean theoretical interpretation of the exclusive rates
requires that we know the corresponding form factors.  (However, $CP$
asymmetries are independent of the form  factors.)  While useful relations
between form factors can be derived from heavy quark symmetry, ultimately
unquenched lattice calculations will be needed for clean theoretical
interpretation of exclusive decays.

Exclusive decays, on the other hand, are experimentally easier to measure. 
There have been recently some very significant theoretical developments towards
understanding the relevant heavy-to-light form factors in the region of
moderate $q^2$ (large recoil).  

\FIGURE{
\epsfig{file=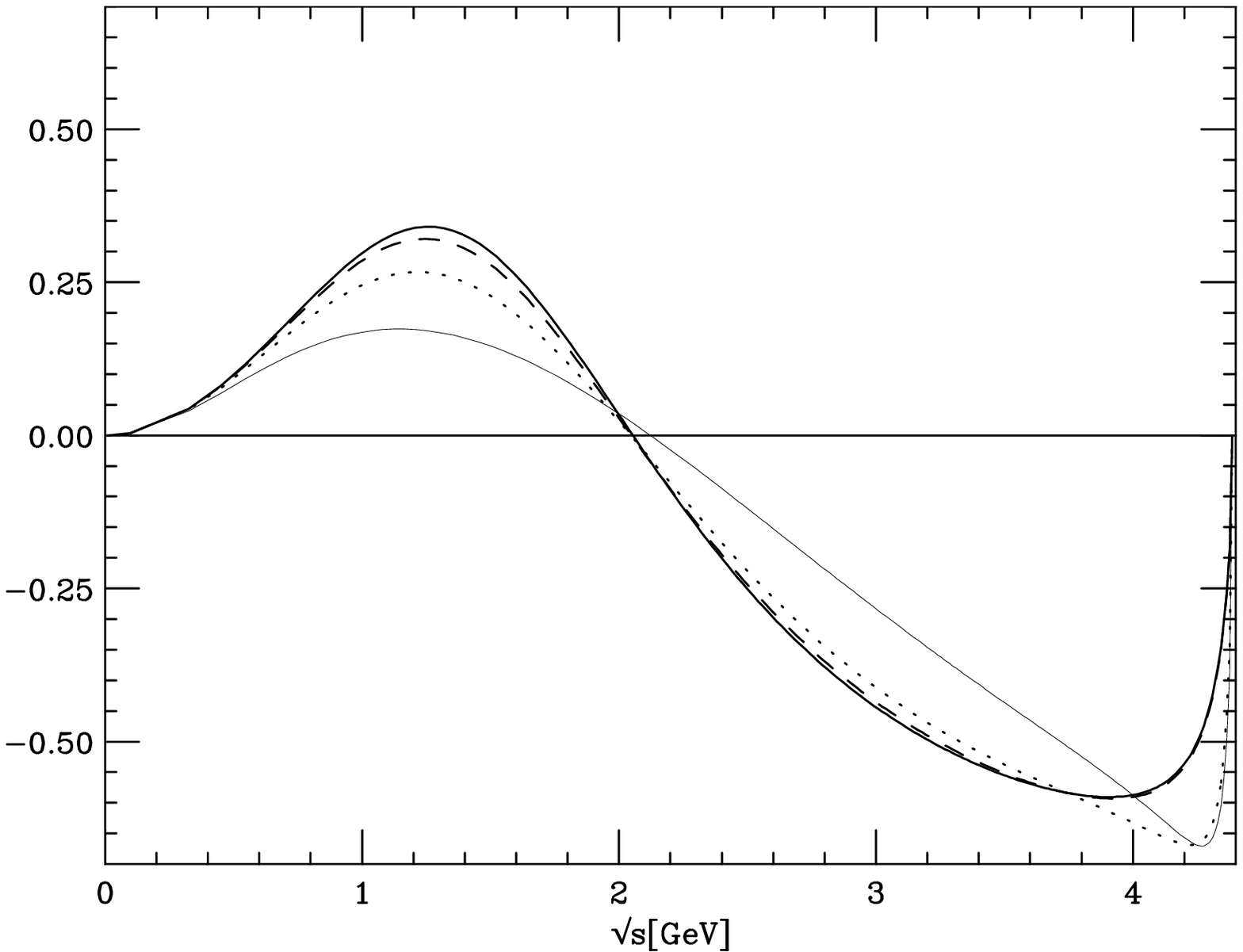, width=.383\textwidth}
\caption{$A_{FB}$ in $B\to K^*\ell^+\ell^-$ in different form factor models 
($s \equiv q^2$)~\cite{burdmanAfb}.}
\label{fig:afb}
}

It was originally observed~\cite{burdmanAfb} that $A_{FB}$, the forward-backward
asymmetry in $B\to K^*\ell^+\ell^-$, vanishes at a value of $q^2$ independent
of form factor models (near $q_0^2 = 4\,\mbox{GeV}^2$ in the SM, see
Fig.~\ref{fig:afb}).  This was shown to follow model independently from the 
large energy limit discussed below~\cite{charles,BFPS}.  One finds the
following implicit equation for $q_0^2$
\begin{equation}
C_9(q_0^2) = - C_7\, {2m_B m_b\over q_0^2}\,  
  \bigg[ 1 + {\cal O}\bigg(\alpha_s, \frac\lqcd{m_b}\bigg) \bigg] .
\end{equation}
The order $\alpha_s$ corrections are calculable~\cite{BF,BFS}, but one cannot
reliably estimate the $\lqcd/E_{K^*}$ terms yet.  Nevertheless, these results
will allow to search for new physics in $A_{FB}$; $C_7$ is known from $B\to
X_s\gamma$, so the zero of $A_{FB}$ determines $C_9$, which is sensitive to new
physics ($C_{7,9}$ are the effective Wilson coefficients often denoted by
$C_{7,9}^{\rm eff}$, and $C_9$ has a mild $q^2$-dependence).

The above simplifications occur because the 7 form factors that parameterize
all $B\to \mbox{vector meson}$ transitions ($B\to K^*\ell^+\ell^-$,
$K^*\gamma$, or $\rho\ell\bar\nu$) can be expressed in terms of only two
functions, $\xi_\perp(E)$ and $\xi_\|(E)$, in the limit where $m_b\to\infty$
and $E_{\rho,K^*} = {\cal O}(m_b)$~\cite{charles}.  In the same limit, the 3
form factors that parameterize decays to pseudoscalar mesons are related to one
function, $\xi_P(E)$.  We are just beginning to see the foundations of these
ideas clarified~\cite{BFPS}, and applications worked out.  E.g., the $B\to
K^*\gamma$ rate can be used to constrain the $B\to \rho\ell\bar\nu$ form
factors relevant for $|V_{ub}|$~\cite{BH}.  The large ${\cal O}(\alpha_s)$
enhancement of $B\to K^*\gamma$ together with the agreement between the
measured rate and the leading order prediction using  light cone sum rules for
the form factor imply that the form factor predictions have sizable errors or
the subleading terms in $\lqcd/E_{\rho,K^*}$ are significant~\cite{BFS,BoBu}. 
How well the theory can describe these processes will test some of the 
ingredients entering factorization in charmless $B$ decays.

\subsection{Semileptonic and rare decays --- Summary}

\begin{itemize}\itemsep 0pt

\item $|V_{cb}|$ is known at the $\lsim5\%$ level; error may become half of
this in the next few years using both inclusive and exclusive measurements. The
inclusive requires precise determination of $m_b$ using various spectra and
tests of duality, the exclusive will rely on the lattice.  

\item Situation for $|V_{ub}|$ may become similar to present $|V_{cb}|$.  For a
precise inclusive measurement the neutrino reconstruction to obtain $q^2$ and
$m_X$ seems crucial (and determining $m_b$ as mentioned above); the exclusive
will require unquenched lattice calculations.

\item Important progress towards understanding exclusive rare decays in the
small $q^2$ regime, $B\to \rho\ell\bar\nu$, $K^{(*)}\gamma$, and
$K^{(*)}\ell^+\ell^-$ below the $\psi$.  This increases the sensitivity to new
physics, and may also test some ingredients entering factorization in charmless
decays.

\end{itemize}\vspace*{-10pt}

\section{Nonleptonic decays}\label{sec:nl}

In this Section I discuss three topics where important developments occurred
recently.  The first is factorization in exclusive hadronic $B$ decays. 
Especially charmless decays are very important for studying $CP$ violation. 
The second is inclusive widths and lifetimes, where OPE based calculations are
possible.  The third is $D-\D0bar$ mixing, where there have also been new
experimental and theoretical results.

\subsection{\boldmath Factorization in exclusive $B$ decays}

\FIGURE{
\epsfig{file=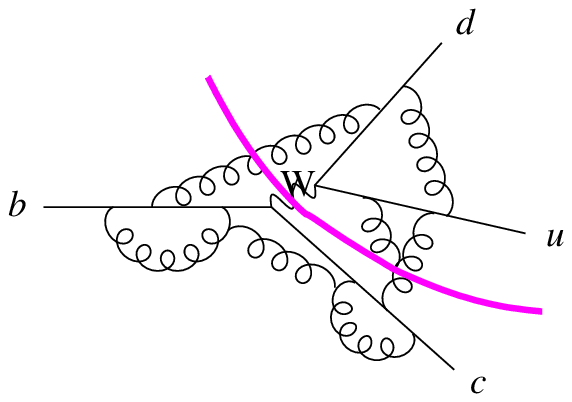, width=0.312\textwidth}
\caption{Sketch of factorization in $\B0bar\to D^{(*)+}\pi^-$ decay.}
\label{fig:factplot}
}

Until recently very little was known model independently about exclusive
nonleptonic $B$ decays.  Crudely speaking, factorization is the hypothesis
that, starting from the effective nonleptonic Hamiltonian, one can estimate
matrix elements of four-quark operators by grouping the quark fields into a
pair that can mediate $B\to M_1$ decay ($M_1$ inherits the ``brown muck" of the
decaying $B$), and another pair that can describe $\mbox{vacuum} \to M_2$
transition.  E.g., in $\B0bar\to D^{(*)+}\pi^-$, this amounts to the assumption
that the contributions of gluons ``parallel" to the $W$ are calculable
perturbatively or suppressed by $\lqcd/m_Q$ (see Fig.~\ref{fig:factplot}).

It has long been known that if $M_1$ is heavy and $M_2$ is light, such as
$\B0bar\to D^{(*)+}\pi^-$, then ``color transparency" may justify
factorization~\cite{BJfact,DGfact,PWfact}.  The physical picture is that the
two quarks forming the $\pi$ must emerge from the short distance process in a
small color dipole state (two fast collinear quarks in a color singlet), and at
the same time the wave function of the brown muck in the $B$ only changes
moderately since the $D$ recoil is small.  Recently it was shown  to
2-loops~\cite{BBNS}, and to all orders in perturbation theory~\cite{BPS}, that
in such decays factorization is the leading result in a systematic expansion in
powers of $\alpha_s(m_Q)$ and $\lqcd/m_Q$.  While the $\alpha_s$ corrections
are calculable, little is known from first principles about those suppressed,
presumably, by powers of $\lqcd/m_b$.  A renormalon analysis suggests that in
$\B0bar\to D^{(*)+}\pi^-$, where the light-cone wave function of $M_2$ (the
$\pi$) is symmetric, nonperturbative corrections are actually suppressed by two
powers~\cite{renormfac}.

It is important to test experimentally how well factorization works, and learn
about the size of power suppressed effects.  The $\langle D^{(*)} | \bar
c_L\gamma^\mu b_L | \B0bar \rangle$ matrix element is measured in semileptonic
$B\to D^{(*)}$ decay, while $\langle X | \bar u_L\gamma^\mu d_L | 0 \rangle$
for $X = \pi, \rho$ is given by the known decay constants.  Thus, in ``color
allowed" decays, such as $\B0bar\to D^{(*)+}\pi^-$ and $D^{(*)+}\rho^-$,
factorization has been observed to work at the $\sim 10\%$ level.  These tests
get really interesting just around this level, since there is another argument
that supports factorization, which is independent of the heavy mass limit.  It
is the large $N_c$ limit ($N_c = 3$ is the number of colors), which implies for
such decays that factorization violation is suppressed by $1/N_c^2$.  The large
$N_c$ argument for factorization is independent of the final state, whereas the
one based on the heavy quark limit predicts that the accuracy depends on the
kinematics of the decay.

One of the predictions of QCD factorization in $B\to D\pi$ is that amplitudes
involving the spectator quark in the $B$ going into the $\pi$ should be power
suppressed~\cite{BBNS}, and therefore,
\begin{equation}
{\cal B}(B\to D^{(*)0} \pi^-) \big/ {\cal B}(B\to D^{(*)+} \pi^-)
  = 1 + {\cal O}(\lqcd/m_Q) \,.
\end{equation}
However, experimentally, this ratio is in the ballpark of 1.8 with errors
around 0.3 for both $D$ and $D^*$ and also for $\pi$ replaced by $\rho$.  This
has been argued to be due to ${\cal O}(\lqcd/m_c)$ corrections, which may be
sizable~\cite{BBNS}.

\TABLE{
\begin{tabular}{ccc}\hline
${\cal B}(B\to D^0 \pi^0)$  &  ${\cal B}(B\to D^{*0} \pi^0)$  &  
  $[\times 10^{-4}]$  \\ \hline\hline
$3.1\pm0.4\pm0.5$  &  $2.7^{+0.8\,+0.5}_{-0.7\,-0.6}$  &
  BELLE~\cite{BELLEcolsup} \\
$2.74^{+0.36}_{-0.32}\pm0.55$  &  $2.20^{+0.59}_{-0.52}\pm0.79$  &
  CLEO~\cite{CLEOcolsup} \\ \hline
\end{tabular}%
\caption{Color suppressed $B\to D^{(*)0} \pi^0$ branching ratios.}
\label{tab:colsup}
}

The first observations of ``color suppressed" $B$ decays, $B\to D^{(*)0}
\pi^0$, were reported at this conference.  The results are summarized in
Table~\ref{tab:colsup}.  These rates are larger than earlier theoretical
expectations (or than the upper bound for $D^0\pi^0$ in the Y2K PDG).  This
data allows, for the first time, to extract the strong phase difference between
the $\Delta I = \frac32$ and $\frac12$ amplitudes from the measured $B\to
D^+\pi^-$, $D^0\pi^-$, and $D^0\pi^0$ rates.  Factorization predicts that this
phase should be power suppressed.  My slides at the conference showed that this
phase was around $24^\circ$ with asymmetric errors around $6^\circ$.  Since
then, several analyses are published with varying conclusions about the meaning
of these results~\cite{strongphase}.  It will be interesting to see what
happens when the experimental errors decrease.

There are many other testable predictions.  E.g., factorization also holds in
$\B0bar\to D^{(*)+} D_s^{(*)-}$ within the (presently sizable) errors, which is
interesting because the heavy $D_s$ meson must come from the $W$
boson~\cite{LR}.  At some level one expects to see deviations from
factorization in this decay which are larger than those in $\B0bar\to D^{(*)+}
\pi^-$.  When the $B\to \pi$ semileptonic form factors and the $\B0bar\to \pi^+
D_s^{(*)-}$ rate will be measured, it will be interesting to compare the
accuracy of factorization with that in $\B0bar\to D^{(*)+}\pi^-$.  In QCD
factorization $\B0bar\to \pi^+ D_s^{(*)-}$ is power suppressed, so corrections
to ``naive factorization" are not subleading in the power counting.  So I would
not trust $|V_{ub}|$ extracted from this rate.

It was also observed that in $B\to D^{(*)}X$, where $X$ is a meson with spin
greater than one or has a small decay constant (such as the $a_0$, $b_1$, etc.,
which can only be created by the weak current due to isospin breaking), the
leading factorizable term vanishes, but there is a calculable ${\cal
O}(\alpha_s)$ contribution~\cite{Gudrun}.  Unfortunately there are also power
suppressed uncalculable corrections, which may be comparable.  Such ideas could
also be useful for $CP$ violation studies in charmless decays, by suppressing
certain tree amplitudes~\cite{Gudrun,LS}.  There may be preliminary evidence
for one such decay, $B\to a_0\pi$~\cite{babara0pi}.

\FIGURE{
\hfill\raisebox{-2.6pt}{\epsfig{file=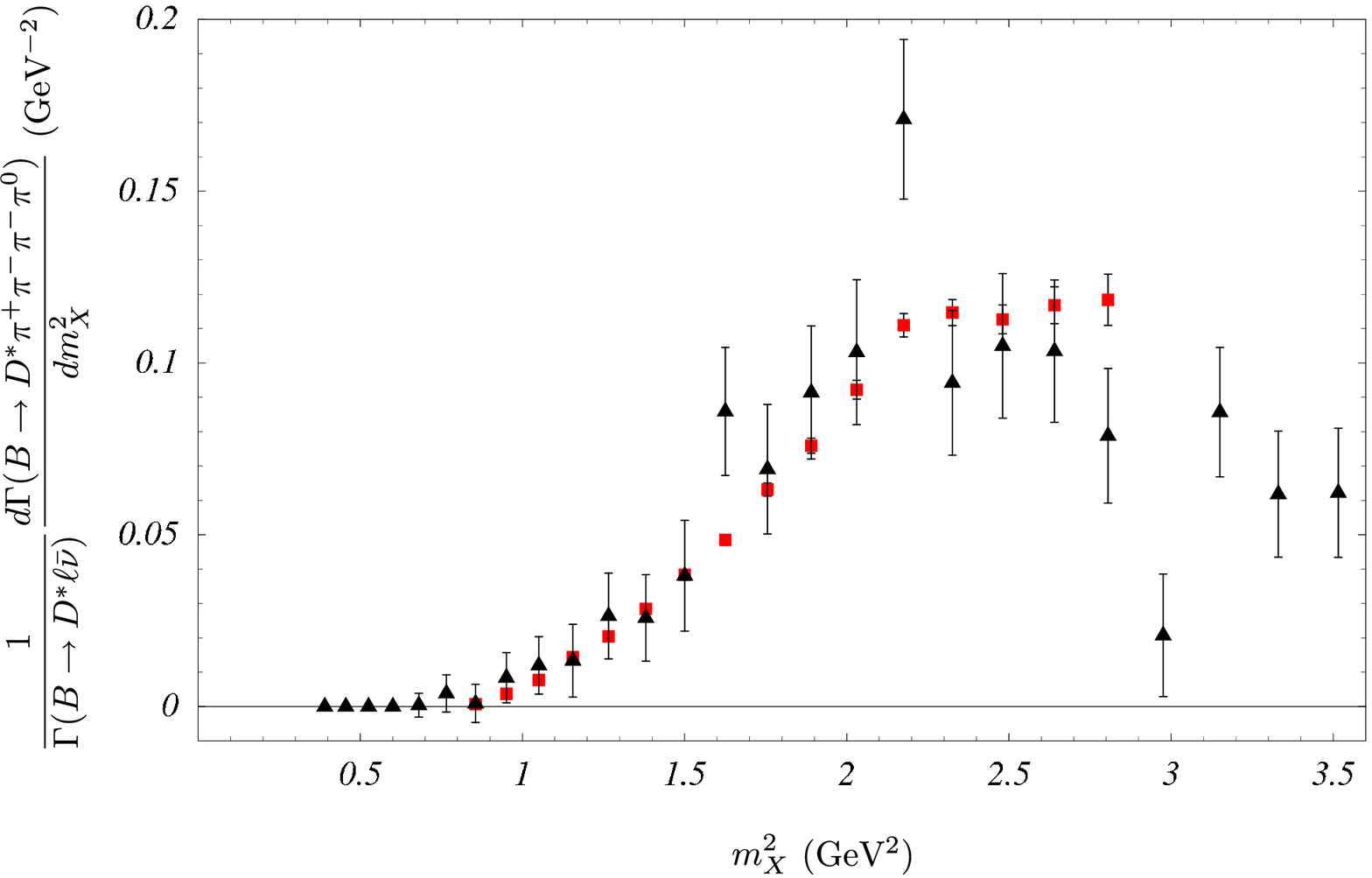, width=0.459\textwidth}}
\hfill\hfill \epsfig{file=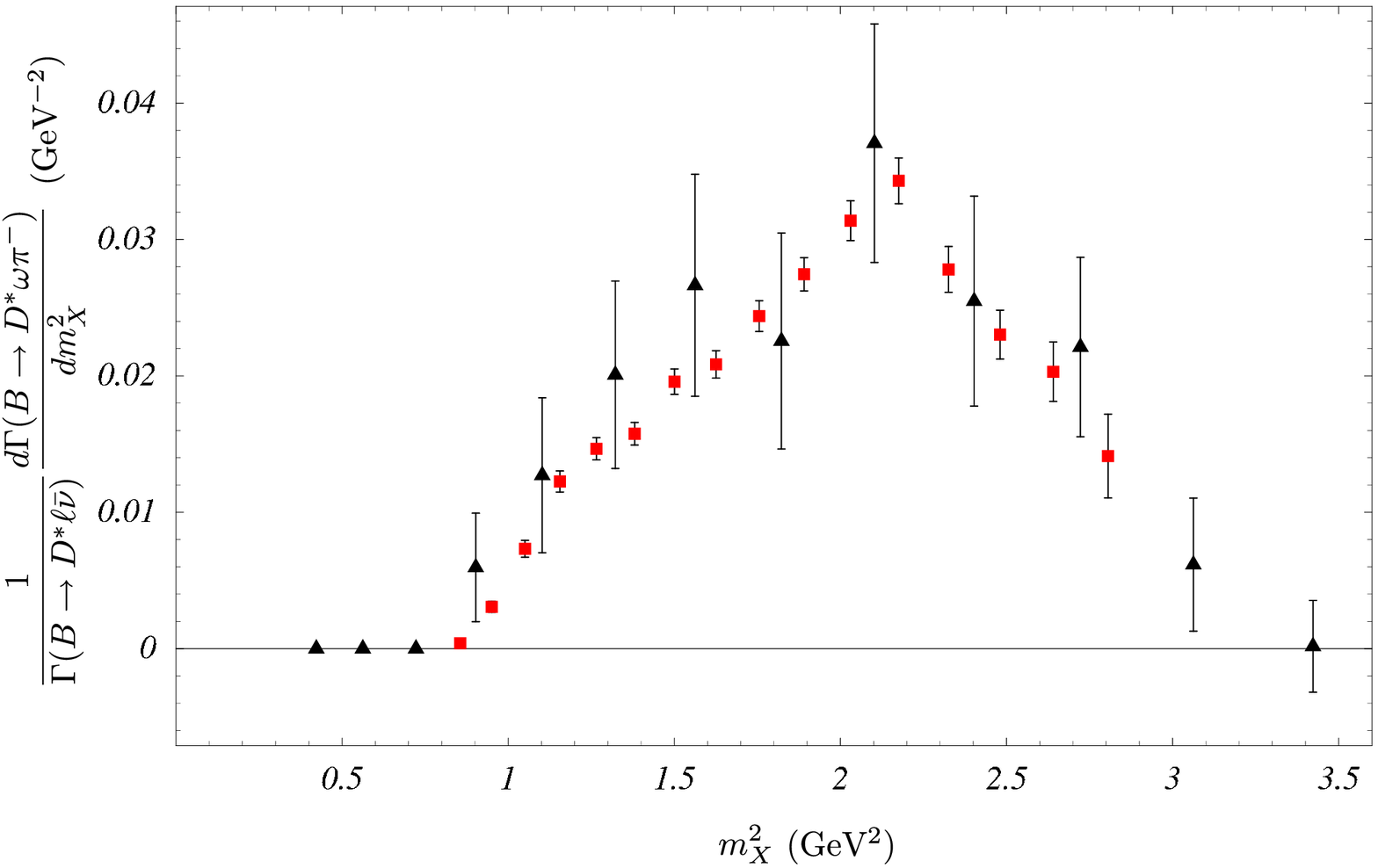, width=0.45\textwidth}
\caption{$\d\Gamma(B\to D^*\pi^+\pi^-\pi^-\pi^0)/\d m_X^2$, where  $m_X$ is the
$\pi^+\pi^-\pi^-\pi^0$ invariant mass (left), and $\d\Gamma(B\to
D^*\omega\pi)/\d m_X^2$, where $m_X$ is the $\omega\pi^-$ invariant mass
(right), normalized to the $B\to D^*\ell\bar\nu$ rate.  Black triangles are $B$
decay data, red squares are the predictions using $\tau$ data~\cite{LLW}.}
\label{fig:4pi}
}

Multi-body $B\to D^{(*)} X$ modes have also been used to study corrections to
factorization~\cite{LLW}.  The advantage compared to two-body channels is that
the accuracy of factorization can be studied for a final state with fixed
particle content, by examining the differential decay rate as a function of the
invariant mass of the light hadronic state $X$ (this was also suggested in
Refs.~\cite{DGfact,ReIs}).  If factorization works primarily due to the large
$N_c$ limit then its accuracy is not expected to decrease as the $X$ invariant
mass, $m_X$, increases.  If factorization is mostly due to perturbative QCD
then there should be corrections which grow with $m_X$.  Combining data for
hadronic $\tau$ decays and semileptonic $B$ decays allows such tests to be made
for a variety of final states.  Fig.~\ref{fig:4pi} shows the comparison of the
$B\to D^*\pi^+\pi^-\pi^-\pi^0$ and $D^*\omega\pi^-$ data~\cite{CLEOD4pi} with
the $\tau$ decay data~\cite{CLEOtau4pi}.  The kinematic range accessible in
$\tau\to 4\pi$ corresponds to $0.4 \lsim m_{4\pi}/E_{4\pi} \lsim 0.7$ in $B\to
4\pi$ decay.  A background to these comparisons is that one or more of the
pions may arise from the $\bar c_L \gamma^\mu b_L$ current creating a
nonresonant $D^* + n\,\pi$ ($1 \leq n \leq 3$) state or a higher $D^{**}$
resonance.  In the $\omega\pi^-$ mode this is very unlikely to be
significant~\cite{LLW}.  In the $\pi^+\pi^-\pi^-\pi^0$ mode such backgrounds
can be constrained by measuring $B\to D^*\pi^+\pi^+\pi^-\pi^-$, since
$\pi^+\pi^+\pi^-\pi^-$ cannot come from the $\bar u_L \gamma^\mu d_L$ current. 
CLEO found ${\cal B}(B\to D^*\pi^+\pi^+\pi^-\pi^-) / {\cal B}(B\to
D^*\pi^+\pi^-\pi^-\pi^0) < 0.13$ at 90\%\,CL in the  $m_X^2 < 2.9\,{\rm GeV}^2$
region~\cite{CLEO4piws}.  With more precise data, observing deviations that
grow with $m_X$ would be evidence that perturbative QCD is an important part of
the success of factorization in $B\to D^*X$.

Calculating $B$ decay amplitudes to charmless two-body final states is
especially important for the study of $CP$ violation.  There are two approaches
to these decays.  BBNS~\cite{bbnslight} assume that Sudakov suppression is not
effective at the $B$ mass scale in the endpoint regions of quark distribution
functions, while Keum {\it et al.}~\cite{keumetal} assume that it is.  They
yield different power counting and often different phenomenological
predictions.  In the former approach the $B\to \pi\ell\bar\nu$ form factors are
nonperturbative functions to be determined from data, while they are calculable
in the latter.  My guess would be that they are not calculable (they would be
if $m_b$ were huge), but it will take time to really decide this using data. 
Predictions for direct $CP$ violation are often smaller in the former than in
the latter approach.  An outstanding open theoretical question is the complete
formulation of power suppressed corrections.  Some of them are known to be
large, e.g., the ``chirally enhanced" terms proportional to $m_K^2/(m_s m_b)$
which are not enhanced by any parameter of QCD in the chiral limit, just
``happen to be" large, and the uncertainty related to controlling the infrared
sensitivity in annihilation contributions.  (See also the discussion of these
issues in Refs.~\cite{HQhf9,MBWlp}.)  It has also been claimed that the effects
of charm loops are bigger than given by perturbation
theory~\cite{charmloops}.

\subsection{\boldmath Inclusive nonleptonic decays, $b$ hadron lifetimes}

Inclusive nonleptonic decay rates of heavy hadrons can also be computed in an
OPE, like inclusive semileptonic rates.  The crucial difference is that the OPE
has to be performed in the physical region, and so lifetime predictions rely on
local duality, whereas inclusive semileptonic rates only rely on global
duality.  Formally, they are expected to have similar accuracy in the $m_b \to
\infty$ limit, but it is quite possible that the scale at which local duality
becomes a good approximation is larger than that for global duality.  It would
not be surprising if the predictions of the OPE work better for semileptonic
than for nonleptonic rates.

\FIGURE{
\epsfig{file=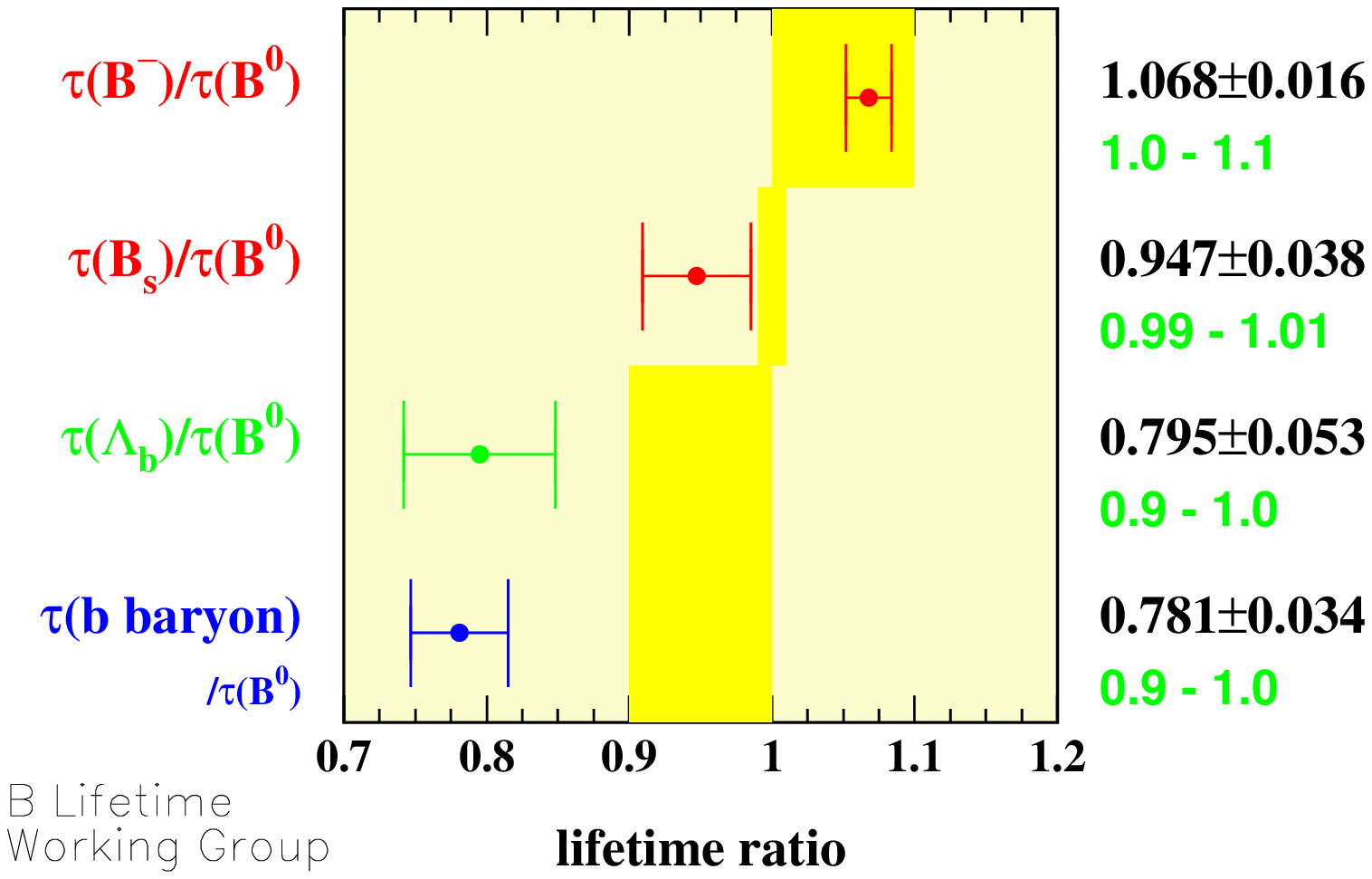, width=0.445\textwidth}
\caption{$b$ hadron lifetime ratios~\cite{Osterberg}.}
\label{fig:lifetimes}
}

The most recent world average $b$ hadron lifetime ratios, together with
theoretical expectations, are shown in
Fig.~\ref{fig:lifetimes}~\cite{Osterberg}.  The lifetime differences are
expected to be dominated by matrix elements of four-quark operators at order
$(\lqcd/m_b)^3$, which have to be determined from lattice
QCD~\cite{MDP,latticelifetime}.  For now, the smallness of $\tau(\Lambda_b)$
remains hard to explain.  However, this is not an indication that semileptonic
widths have similar theoretical uncertainties.  The semileptonic width are in
fact consistent, ${\cal B}(\Lambda_b\to X\ell\bar\nu) / \tau(\Lambda_b) \simeq 
{\cal B}(B\to X\ell\bar\nu) / \tau(B)$, within the $\sim15\%$ error of the
present experimental data~\cite{Palla}.

The assumption in the OPE calculation of nonleptonic widths related to local
duality has been questioned recently.  In the 't~Hooft Model (two dimensional
QCD) it was found numerically that the widths of a heavy meson and a heavy
quark differ by order $\lqcd/m_Q$, and one needs to do an (unphysical) smearing
over $m_Q$ to reduce the discrepancy to $\lqcd^2/m_Q^2$~\cite{benincl}.

\subsection{\boldmath $D^0-\D0bar$ mixing}

The $D^0$ system is unique among the neutral mesons in that it is the only one
whose mixing proceeds via intermediate states with down-type quarks. 
$D^0-\D0bar$ mixing is a sensitive probe of new physics, because the SM
prediction for $x \equiv \Delta M_D/\Gamma_D$, $y \equiv \Delta
\Gamma_D/2\Gamma_D$, and the $CP$ violating phase in the mixing, $\phi$, are
very small.  While $y$ is expected to be dominated by SM processes, $x$ and
$\phi$ could be significantly enhanced by new physics.

\TABLE{
\begin{tabular}{c|ccc} \hline
ratio  &  4-quark  &  6-quark  &  8-quark \\ \hline\hline
$\displaystyle {\Delta M\over\Delta M_{\rm box}}$  &  1  
  &  $\displaystyle {\Lambda^2 \over m_s m_c}$  
  &  $\displaystyle {\alpha_s\over4\pi}\, {\Lambda^4\over m_s^2 m_c^2}$ 
  \\ \hline
$\displaystyle {\Delta \Gamma\over \Delta M}$  
  &  $\displaystyle {m_s^2\over m_c^2}$
  &  $\displaystyle {\alpha_s\over 4\pi}$  
  &  $\displaystyle {\alpha_s\over 4\pi}\, \beta_0$ \\ \hline
\end{tabular} \vspace{4pt}
\caption{$\Delta M$ and $\Delta\Gamma$ in the OPE ($\Lambda \sim  1\,$GeV).}
\label{tab:opetable}
}

$D^0$ mixing is very slow in the SM, because the third generation plays a
negligible role due to the smallness of $|V_{ub} V_{cb}|$, the GIM cancellation
is very effective due to the smallness of $m_b/m_W$, and the mixing is also
suppressed by $SU(3)$ breaking.   $x$ and $y$ are hard to estimate reliably
because the charm quark is neither heavy enough to trust the ``inclusive"
approach based on the OPE, nor light enough to trust the ``exclusive" approach
which sums over intermediate hadronic states.  The short distance box diagram
contributes $x_{\rm box} \sim {\rm few}\times 10^{-5}$ since it is suppressed
by $m_s^4/(m_W^2 m_c^2)$, and $y_{\rm box} \sim {\rm few}\times 10^{-7}$ since
it has an additional $m_s^2/m_c^2$ helicity suppression.  Higher order terms in
the OPE are very important, because they are suppressed by fewer powers of
$m_s$ (see Table~\ref{tab:opetable})~\cite{Ge92,BiUr,FGLP}.  With large
uncertainties due to the hadronic matrix elements, most estimates yield $x,y
\lsim 10^{-3}$.

\TABLE{
\begin{tabular}{rc} \hline
Value of $y_{CP}$  &  Experiment \\ \hline\hline
$0.8 \pm 3.1 \%$  &  E791~\cite{e791y} \\
$3.4 \pm 1.6 \%$   &  FOCUS~\cite{focusy} \\
$-1.1 \pm 2.9 \%$  &  CLEO~\cite{cleoy} \\
$-0.5 \pm 1.3 \%$   &  BELLE~\cite{belley} \\
$-1.0 \pm 2.8 \%$  &  BABAR~\cite{babary} \\ \hline
\end{tabular}
\caption{$y_{CP}$ measurements.}
\label{tab:yCP}
}

There are three types of experiments which measure $x$ and $y$.  Each is
actually sensitive to a combination of $x$ and $y$, rather than to either
quantity directly.  First, the $D^0$ lifetime difference to $CP$ even and $CP$
odd final states can be measured by comparing the lifetimes to a flavor and a
$CP$ eigenstate.  To leading order,
\beq\label{ycp}
y_{CP} = {\tau(D \to \pi^+ K^-) \over \tau(D \to K^+ K^-)} - 1
     = y\cos\phi - x \sin\phi\, \frac{A_m}2\,,
\eeq

\FIGURE{
\epsfig{file=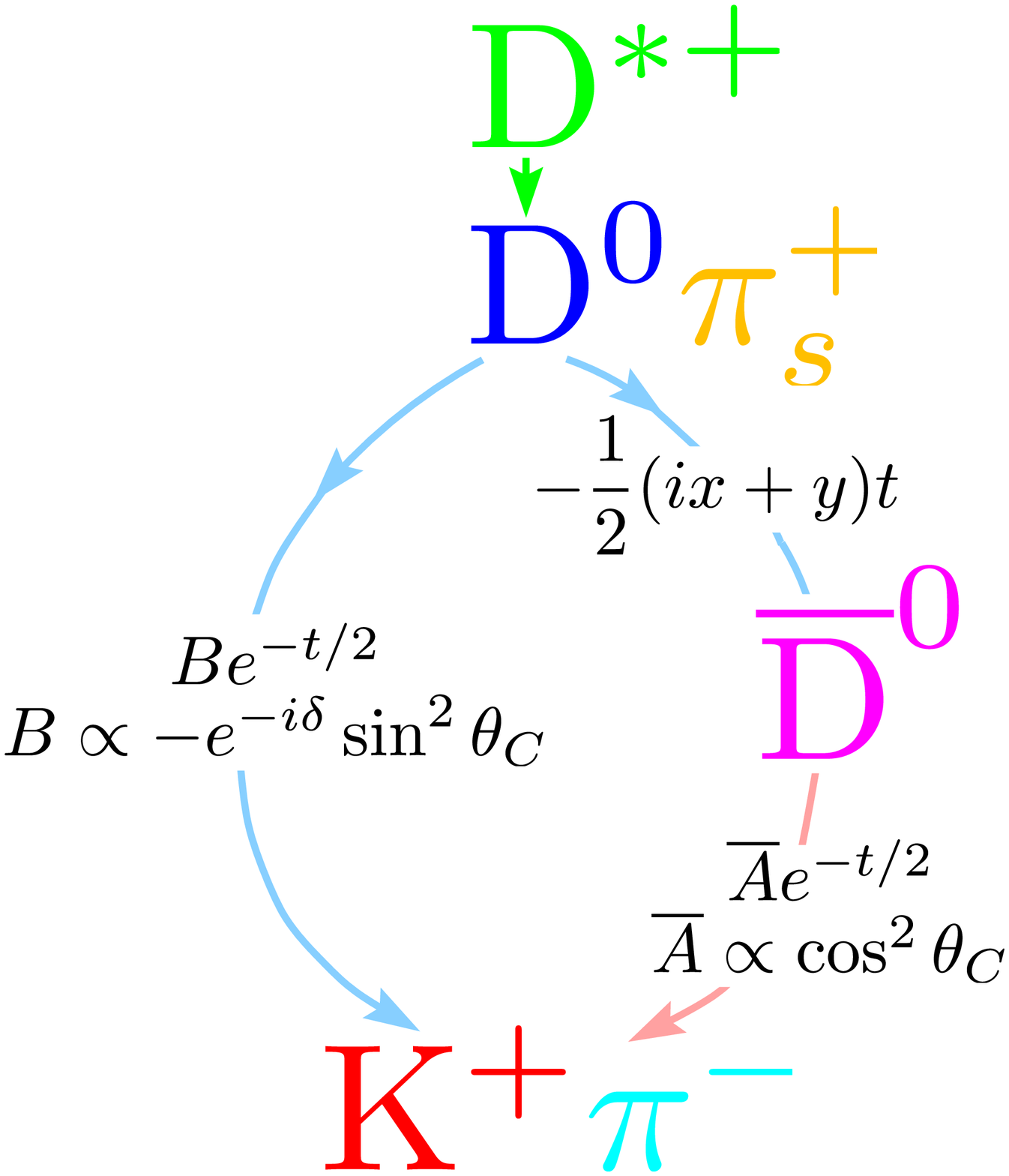, width=0.24\textwidth}
\vspace*{-1.5cm}
\caption{$D^0\to K^+ \pi^-$.}
\label{fig:dmix}
}\noindent
where $A_m = |q/p|^2-1$, which is very small in the SM.  The present data in
Table~\ref{tab:yCP} yield a world average $y_{CP} \simeq 0.65 \pm 0.85
\%$.\break
Second, the time dependence of doubly Cabibbo suppressed decays, such as
$D^0\to K^+ \pi^-$~\cite{Godang:2000yd}, is sensitive to the three quantities
\beq\label{xyprime}
(x \cos\delta + y \sin\delta) \cos\phi\,, \qquad
  (y \cos\delta - x \sin\delta) \sin\phi\,, \qquad
  x^2 + y^2\,,
\eeq
where $\delta$ is the strong phase between the Cabibbo allowed and doubly
Cabibbo suppressed amplitudes (see Fig.~\ref{fig:dmix}).  A similar study for
$D^0\to K^-\pi^+\pi^0$ would be valuable, with the strong phase difference
extracted simultaneously from the Dalitz plot analysis~\cite{brand}. Third, one
can search for $D$ mixing in semileptonic decays~\cite{DmixSL}, which is
sensitive to $x^2+y^2$.

Although $y$ is expected to be determined by Standard Model processes, its
value affects significantly the sensitivity to new physics~\cite{BGLNP}.  If
$y$ is larger or much larger than $x$, then the observable $CP$ violation in
$D^0$ mixing is necessarily small, even if new physics dominates $x$.  A recent
estimate of $y$ calculated $SU(3)$ breaking in phase space differences, and
found that $y \sim 1\%$ can easily be accommodated in the SM~\cite{FGLP}. 
Final states in $D^0$ decay can be decomposed in representations of $SU(3)$.
The cancellation between decays to members of a given representation can be
significantly violated because the final states containing larger number of
strange hadrons have smaller phase space, or can even be completely forbidden. 
Such effects might enhance $y$ more significantly than they affect $x$.

Therefore, searching for new physics and $CP$ violation in $D^0-\D0bar$ mixing
should aim at precise measurements of both $x$ and $y$, and at more complicated
analyses involving the extraction of the strong phase in the time dependence of
doubly  Cabibbo suppressed decays.

\subsection{Nonleptonic decays --- Summary}

\begin{itemize}

\item In nonleptonic $B\to D^{(*)} X$ decay, where $X$ is a low mass hadronic
state, factorization has been established in the heavy quark limit, at leading
order in $\lqcd/m_Q$. 

\item Flood of new and more precise data will allow many tests of factorization
and tell us the significance of unknown power suppressed terms, hopefully also
in charmless decays.

\item In the $D$ system the only unambiguous signal of new physics is $CP$
violation; observation of a large $\Delta m_D$ can only be a clear sign if
$\Delta\Gamma_D$ is smaller, so crucial to measure both.

\end{itemize}\vspace*{-10pt}

\section{Conclusions}\label{sec:concl}

I was not supposed to talk about $CP$ violation, but I had no chance to
succeed, because in order to test the Standard Model in flavor physics all
possible clean measurements which give model independent information on short
distance parameters are very important, whether $CP$ violating or conserving. 

With the recent fairly precise measurement of $\sin2\beta$ and other data, the
CKM contributions to flavor physics and $CP$ violation are likely to be the
dominant ones.  The next goal is not simply to measure $\rho$ and $\eta$, or
$\alpha$ and $\gamma$, but to probe the flavor sector of the SM until it
breaks.  This can be hoped to be achieved in $B$ decays by overconstraining
measurements of the unitarity triangle.  Measurements which are redundant in
the SM but sensitive to different short distance physics are also very
important, since correlations may give information on the new physics we are
encountering (e.g., comparing $\Delta m_s/\Delta m_d$ with ${\cal B}(B\to
X_s\ell^+\ell^-) / {\cal B}(B\to X_d\ell^+\ell^-)$ is not ``just another way"
to measure $|V_{ts}/V_{td}|$).

In many cases hadronic uncertainties are significant and hard to quantify.  
The sensitivity to new physics and the accuracy with which the SM can be tested
will depend on our ability to disentangle the short distance physics from
nonperturbative effects of hadronization.   While we all want smaller errors,
$\epsilon'_K$ reminds us to be conservative with theoretical
uncertainties.  One theoretically clean measurement is worth ten dirty ones. 
But it does change with time what is theoretically clean, and I hope to have
conveyed that there are significant recent developments towards understanding
the hadronic physics crucial both for standard model measurements and for
searches for new physics. For example, (i) for the determination of $|V_{ub}|$
from inclusive $B$ decay; (ii) for understanding exclusive rare decay form
factors at small $q^2$; and (iii) for establishing factorization in certain
nonleptonic decays.

In testing the SM and searching for new physics, our understanding of CKM
parameters and hadronic physics will have to improve in parallel, since except
for a few clean cases (like $\sin2\beta$) the theoretical uncertainties can be
reduced by doing several measurements, or by learning from comparisons with
data how accurate certain theoretical assumptions are.  In some cases data will
help to constrain or get rid of nasty things hard to know model independently
from theory (e.g., excited state contributions to certain processes).  

With the recent spectacular start of the $B$ factories an exciting era in
flavor physics has begun.  The precise measurements of $\sin2\beta$ together
with the sides of the unitarity triangle, $|V_{ub}/V_{cb}|$ at the $e^+e^-$ $B$
factories and $|V_{td}/V_{ts}|$ at the Tevatron, will allow to observe small
deviations from the Standard Model.  The large statistics will allow the study
of rare decays and to improve sensitivity to observables which vanish in the SM
(e.g., certain $CP$ asymmetries); these measurements have individually the
potential to discover physics beyond the SM.  If new physics is seen, then a
broad set of measurements at both $e^+ e^-$ and hadronic $B$ factories and
$K\to \pi \nu\bar\nu$ may allow to discriminate between various scenarios. 
This is a vibrant theoretical and experimental program, and I think the most
concise summary of the status of the field is:

\bigskip
\hfill\parbox{.6\textwidth}{\small
``This is not the end. It is not even the beginning of the end. \\
But it is, perhaps, the end of the beginning."\\
\hspace*{1cm}\hfill W. Churchill~ (Nov.\ 10, 1942)}

\acknowledgments

It is a pleasure to acknowledge numerous discussions with Adam Falk, Mike Luke,
Helen Quinn, and Mark Wise, and many colleagues for help with the recent data,
especially Karl Ecklund, Gautier Hamel de Monchenault, and Vivek Sharma.  I
would like to thank the organizers for the invitation to present this talk and
for their support.  This work was supported in part by the Director, Office of
Science, Office of High Energy and Nuclear Physics, Division of High Energy
Physics, of the U.S.\ Department of Energy under Contract DE-AC03-76SF00098.

\end{document}